\documentclass[twocolumn,aps,pra,amsfonts,amsmath,amssymb,floatfix,nofootinbib]{revtex4}

\usepackage{graphicx}
\usepackage{subfigure}
\usepackage[]{amsfonts, amsmath, amssymb, amstext, latexsym, float}
\usepackage{comment}

\begin{document}
\title{Full analysis of multi-photon pair effects 
in spontaneous parametric down conversion based 
photonic quantum information processing}

%
\author{Masahiro Takeoka, Rui-Bo Jin, and Masahide Sasaki
\\ {\it National Institute of Information and Communications Technology,
Koganei, Tokyo 184-8795, Japan}
}

\begin{abstract}
In spontaneous parametric down conversion (SPDC) based quantum information
processing (QIP) experiments, there is a tradeoff between the coincide count
rates (i.e. the pumping power of the SPDC), which limits the rate of 
the protocol, and the visibility of the quantum
interference, which limits the quality of the protocol. 
This tradeoff is mainly caused by the multi-photon pair emissions 
from the SPDCs. In theory, the problem is how to model the 
experiments without truncating these multi-photon emissions 
while including practical imperfections.

In this paper, we establish a method to theoretically simulate
SPDC based QIPs which fully incorporates the effect of multi-photon
emissions and various practical imperfections.
The key ingredient in our method is
the application of the characteristic function formalism which has been used
in continuous variable QIPs. We apply our method to three examples, 
the Hong-Ou-Mandel interference and the Einstein-Podolsky-Rosen 
interference experiments, and the concatenated entanglement swapping 
protocol. 
For the first two examples, we show that our theoretical results 
quantitatively agree with the recent experimental results. 
Also we provide the closed expressions for these the interference 
visibilities with the full multi-photon components and various imperfections. 
For the last example, we provide the general theoretical form of 
the concatenated entanglement swapping protocol in our method and 
show the numerical results up to 5 concatenations. 
Our method requires only a small computation resource (few minutes by 
a commercially available computer) 
which was not possible by the previous theoretical approach. 
Our method will have applications in a wide range of SPDC based QIP protocols 
with high accuracy and a reasonable computation resource. 
\end{abstract}



\maketitle

\section{Introduction}

Spontaneous parametric down conversion (SPDC) is one of the most standard tools
in photonic quantum information processing (QIP) e.g. 
quantum key distribution, quantum teleportation, quantum repeaters, 
and linear optics quantum computation \cite{Pan2012}. 
Toward implementing higher rate entanglement-based QKD or
larger scale QIP protocols, it is important to increase
the photon-pair generation rate from the SPDC source such that
it provides reasonable coincidence counts of photons in multiple detectors.
Though this is possible by simply increasing the pump power into
the SPDC crystals, it simultaneously degrades the quantum interference
visibility due to the unwanted multi-photon emissions.
Therefore, on one hand, it is an important experimental topic how to reduce
the effect of multi-photon emissions while keeping the higher generation
rates. 
The experimental progress in this direction has been reported recently
\cite{Kuzucu2008, Krischek2010, Ma2011, Broome2011, Jin2014SR}.

On the other hand, in theory, it is desirable to establish
a method which fully incorporates the multi-photon emissions
and various practical imperfections of the experiment, and is also
able to simulate various QIP applications with complicated 
optical circuits. 
The quantum state generated into the signal and idler modes
from an SPDC source is described by
a two-mode squeezed vacuum (TMSV):
\begin{equation}
\label{eq:TMSV}
|\psi\rangle_{SI} = \sqrt{1-\lambda^2} \sum_{n=0}^\infty
\lambda^n |n\rangle_S |n\rangle_I ,
\end{equation}
where $|n\rangle$ represents an $n$-photon state, 
and $\lambda$ is the squeezing parameter. 
Obviously it includes infinitely higher order photons that contribute 
to the multi-photon emissions.
Also, to simulate experiments precisely, one has to take into account 
various imperfections such as losses in channels and detectors, 
dark counts of detectors, mode mismatch between the pulses, and so on.

The theory to describe multi-photon emissions have been 
investigated in various literatures 
\cite{Ou1999,Riedmatten2004,Scarani2005,Scherer2009,Wieczorek2009,Takesue2010,
Broome2011,Jennewein2011,Sekatski2012,Khalique2013,Jin2014SR}. 
A major approach is to calculate the evolution 
of the state vector of the TMSVs in the protocol 
\cite{Ou1999,Riedmatten2004,Scarani2005,Scherer2009,Wieczorek2009,Takesue2010,
Broome2011,Jennewein2011,Khalique2013,Jin2014SR}.
Although the approach is straightforward and useful to 
see the physical insight of multi-photon effects, one of its drawback 
is that one has to truncate the higher photon number 
\cite{Ou1999,Riedmatten2004,Scarani2005} which is not appropriate 
for the higher power pumping. 
In principle, it is possible to include (even infinitely) higher order 
photons in analytical forms. However, the problem is that its mathematical 
expression often becomes complicated even for relatively simple setup  
using only one or two SPDCs like \cite{Scherer2009,Jin2014SR}. 
This is more problematic for a larger scale QIP protocol. 
That is, its numerical simulation often require a huge computational resource 
even truncating the higher order of photons. This severely limits the ability 
to estimate the practical performance of such a protocol. 
For example, in \cite{Khalique2013}, the concatenated operation of 
entanglement swapping is theoretically investigated where the authors 
showed precise but highly complicated mathematical expression of 
the states and detection probabilities in the protocol 
and performed its numerical simulation for the concatenation 
of 3 swappings, which involves 16 optical modes. The simulation was 
performed by a parallel programing on a super computer meaning that 
it requires a huge computational resource even for that scale of the protocol. 
Therefore to extend the analysis for larger SPDC based QIP networks, 
it is desirable to find an alternative way of calculating 
such a problem more systematically, simply, and 
with less computational resources.

As a related work to the above, the authors in \cite{Sekatski2012} developed 
a precise mathematical model of nonideal photon detectors  and then 
derived analytical formulae of several parameters for the experiments 
with one SPDC source, including the Hong-Ou-Mandel (HOM) 
interference and the Einstein-Podolsky-Rosen (EPR) interference 
visibilities where multi-photon pair effects and detector 
imperfections are successfully incorporated. 
Although their formulae are useful for the practical estimation of 
these experiments, it is not fully clear if the approach is easily extendible 
to more complicated protocols such as the concatenated entanglement swapping 
discussed in \cite{Khalique2013}.

In this paper, we propose yet another approach based on 
the phase space representation of quantum optics 
(see \cite{BarnettRadmore} for example) 
which is often used in continuous variable-QIP (CV-QIP) \cite{Braunstein2005}. 
Our method can compute the SPDC based QIP experiments 
by systematically including the multi-photon effects, detector imperfections, 
and moreover, other practical imperfections, such as mode mismatching 
between the pulses from SPDC sources that has not been explicitly considered 
in previous analyses. 
By applying our method to two examples, the HOM interference 
and the EPR interference experiments we show that our method can 
simulate the recent experimental results in \cite{Jin2014OE,Jin2014SR} 
with a quantitative agreement. 
Also we are able to derive closed forms of these visibilities 
that could be handy tools to estimate the effects of multi-photon emissions 
and various imperfections in these experiments.

Moreover, our method can simulate even larger scale SPDC based QIPs 
with a reasonable computational resource. 
As an example, we consider the concatenated entanglement swapping (CES) 
protocol discussed in \cite{Khalique2013} and demonstrate a drastic 
decrease of the required computational resource by our method.
For example, the CES with 3 swappings can be simulated by only 
a 10 second use of a commercial computer and even for 5 swappings, 
it requires around 2 minutes with the same computer. 
This allows us to estimate the optimal number of concatenation for 
a given long-distance channel, for example, a 1000 km optical fiber. 
We believe that our method could be a powerful tool not only for estimating 
the already known protocols mentioned above but for calculating 
the performance of various future QIP protocols based on SPDCs.

The paper is organized as follows. 
In Sec.~\ref{sec:ch_function}, we describe our approach. 
It uses the characteristic function formalism 
which is one of the phase space representation of quantum states and 
operations. We first overview why this approach is 
beneficial for the problems and then review basic definitions and notations
about the characteristic function formalism. We also provide the recipe 
of treating SPDC sources, linear optics, detectors, and various imperfections 
by this formalism. 
In Sec.~\ref{sec:applications}, we apply our method into 
three QIP examples, the HOM interference, the EPR interference, 
and the CES protocol. 
Also we discuss a general computational complexity. 
Note that our method is not generally efficient in the sense that  
the complexity exponentially grows with the size of the system. 
However, as demonstrated in this section, it is still a powerful tool 
for simulating a relatively large size QIP such as the CES protocol 
with a small computational resource. 
Section~\ref{sec:conclusion} concludes the paper.

\section{Characteristic function based approach}
\label{sec:ch_function}

In this section, 
we review the characteristic function formalism and provide the recipe 
to treat the SPDC based QIPs with this formalism. 
Characteristic function and its Fourier transform, Wigner function, 
are typical ways to represent quantum states in phase space and 
are often used in optical CV-QIPs, in particular, for 
{\it Gaussian state and operation}, where the former 
is a quantum state whose characteristic function is a Gaussian function 
and the latter is a quantum operation transforming Gaussian state to 
other Gaussian state \cite{Weedbrook2012}. 
It is known that quantum system consisting of Gaussian states and operations 
are described by calculating only the covariance matrix of 
the Gaussian function which is known to be 
efficiently simulated by classical computation \cite{Bartlett2002} 
(which also means the impossibility of constructing a universal 
quantum computer by only these means).

Typically, the practical SPDC based QIPs consist of SPDC sources, 
linear optics, photon detectors, and imperfections that can be modeled 
by linear operations (note that some of quantum memories can also be 
described by linear operation \cite{Guha2014}). 
The SPDC source, i.e. TMSV, is a typical Gaussian state and 
any linear optics and linear imperfections, that includes 
most of the practical imperfections, are Gaussian operations. 
Therefore, we can efficiently calculate the state evolution 
in the system just before the detection step.

The last part of the experiments, photon detection, 
is non-Gaussian operation. However, our Gaussian approach is still useful 
if we consider the on-off detectors (also called the threshold detectors) 
rather than the photon number resolving detectors. 
On-off detector discriminates only zero or non-zero photons and widely 
used in the SPDC based QIP experiments. 
Such a detector is mathematically described by
two operators, one is a vacuum (zero photons) and the other is
an identity operator minus vacuum (non-zero photons).
Since vacuum is also a Gaussian state, 
one can compute the full process of the protocols 
via the Gaussian function based characteristic function formalism 
(this fact has been recognized in CV-QIPs, 
see \cite{Olivares2005, Takeoka2008, Weedbrook2012} for example). 
In the following, we describe the detailed definitions and notations.

\subsection{Characteristic function}
Let us consider an $n$-mode bosonic system associated with
an infinite-dimensional Hilbert space $\mathcal{H}^{\otimes n}$
and $n$ pairs of annihilation and creation operators,
$\{\hat{a}_i,\, \hat{a}_i^\dagger\}_{i=1,\cdots,n}$, respectively,
which satisfy the commutation relations
\begin{equation}
\label{eq:commutation_relation_a_a^dagger}
[ \hat{a}_i , \hat{a}_j^\dagger ] = \delta_{ij} .
\end{equation}
From these, one may construct the quadrature field operators:
\begin{eqnarray}
\label{eq:quadrature_x}
\hat{x}_i = \frac{1}{\sqrt{2}} ( \hat{a}_i^\dagger + \hat{a}_i ),
\quad
\hat{p}_i = \frac{i}{\sqrt{2}} ( \hat{a}_i^\dagger - \hat{a}_i ).
\end{eqnarray}
It is easy to verify that the commutation relations now translate to $[\hat{x}_i, \hat{p}_j]=i \delta_{ij}$. In the $n$-mode bosonic system, a quantum state with density operator $\hat{\rho}$ is described by its characteristic function
\begin{equation}
\label{eq:characteristic_function}
\chi (x) = {\rm Tr} \left[ \hat{\rho} \hat{\mathcal{W}} (x) \right] ,
\end{equation}
where,
\begin{equation}
\label{eq:Weyl_op}
\hat{\mathcal{W}} (x) = \exp \left[ - i x^T \hat{R} \right] ,
\end{equation}
is a Weyl operator,
$\hat{R}=[ \hat{x}_1, \dots, \hat{x}_n , \hat{p}_1, \dots, \hat{p}_n]^T$
is a $2n$ vector consisting of quadrature operators,
and $x =[ x_1, \cdots , x_{2n} ]$
is a $2n$ real vector.

\subsection{Gaussian states}
{\bf Definition}.
The Gaussian state is defined as the quantum state whose characteristic 
function is given by a Gaussian function: 
\begin{equation}
\label{eq:chi_gaussian}
\chi (x) = \exp\left[ - \frac{1}{4} x^T \gamma x - i d^T x \right] ,
\end{equation}
where $\gamma$ is a $2n \times 2n$ matrix and $d$ is a $2n$ vector 
called the covariance matrix and the displacement vector, respectively.
For example, coherent state $|\alpha\rangle$ is a Gaussian state.
Its characteristic function and displacement vector are described by 
the Gaussian form in Eq.~(\ref{eq:chi_gaussian}) with 
\begin{equation}
\label{eq:gamma_d_coh}
\gamma^{\rm coh} = I,
\quad d = \sqrt{2}
\left[
\begin{array}{c}
{\rm Re}\alpha \\
{\rm Im}\alpha
\end{array}
\right]
\end{equation}
where $I$ is a 2-by-2 identity matrix 
(thus the vacuum state is simply given by $\{\gamma=I, d=0\}$).
The most important Gaussian state in this paper is the TMSV.
Its covariance matrix is given by
\begin{equation}
\label{eq:TMSV}
\gamma^{\rm TMSV}(\mu) = \left[
\begin{array}{cc}
\gamma^+(\mu) & {\bf 0}\\
{\bf 0} & \gamma^-(\mu)
\end{array}
\right] ,
\end{equation}
where
\begin{equation}
\label{eq:TMSV2}
\gamma^\pm(\mu) = \left[
\begin{array}{cc}
2\mu +1 & \pm 2\sqrt{\mu(\mu+1)} \\
\pm 2\sqrt{\mu(\mu+1)} & 2\mu +1 \\
\end{array}
\right] ,
\end{equation}
and $\mu = \lambda^2 /(1-\lambda^2)$ while $d=0$. 
It is worth to note that $\mu$ corresponds to the average photon number 
per mode, i.e. $\mu=\langle\psi| \hat{n} \otimes \hat{I} |\psi\rangle =
\langle\psi| \hat{I}\otimes\hat{n} |\psi\rangle$, 
where $\hat{I}$ and $\hat{n}$ are the identity and photon number 
operators, respectively.

{\bf Partial trace of Gaussian states}. 
The covariance matrix of the reduced state after partial trace
is simply given by the submatrix corresponding to
the remained system.
For example, the covariance matrix of the reduced state of the TMSV
$\rho_{\rm S} = {\rm Tr}_I [|\psi\rangle\langle\psi|_{SI}]$ is
\begin{equation}
\label{eq:partial_trace}
\gamma^{\rm S} = \left[
\begin{array}{cc}
2\mu+1 & 0 \\
0 & 2\mu+1
\end{array}
\right] ,
\end{equation}
which corresponds to the covariance matrix of
a thermal state with average photon number $\mu$: 
\begin{equation}
\label{eq:thermal_state}
\hat{\rho}_{\rm th} = 
\sum_{n=0}^{\infty} \frac{\mu^n}{(\mu+1)^{n+1}} |n \rangle\langle n|. 
\end{equation}

\subsection{Gaussian unitary operations}
Gaussian unitary operation is defined as the unitary operation
that transforms Gaussian states to other Gaussian states.
Any Gaussian unitary operation acting on a Gaussian state
can be described by symplectic transformations
of the covariance matrix and the displacement vector of the state:
\begin{equation}
\label{eq:symplectic}
\gamma \to S^T \gamma S, \quad d \to S^T d,
\end{equation}
where $S$ is a symplectic matrix corresponding to the Gaussian unitary
operation and $T$ is a transpose operation. 
For any covariance matrix, there exists a symplectic
transformation that diagonalizes the covariance matrix
(symplectic diagonalization). If the unitary operation includes
only linear optical process (beam splitters and phase shifts),
then $S^T = S^{-1}$ and such a matrix $S$ is called
an orthogonal symplectic matrix.
The explicit expression of the symplectic matrix for 
phase shifting and beam splitting are given below.

{\bf Phase shift}:
\begin{equation}
\label{eq:phase_shift}
R(\phi) = \left[
\begin{array}{cc}
\cos\phi & \sin\phi \\
-\sin\phi & \cos\phi
\end{array}
\right] .
\end{equation}

{\bf Beam splitter on mode $A$ and $B$}:
\begin{equation}
\label{eq:beam_splitter}
S^{t}_{AB}
= \left[
\begin{array}{cccc}
\sqrt{t} & \sqrt{1-t} & 0 & 0 \\
-\sqrt{1-t} & \sqrt{t} & 0 & 0 \\
0 & 0 & \sqrt{t} & \sqrt{1-t} \\
0 & 0 & -\sqrt{1-t} & \sqrt{t}
\end{array}
\right] ,
\end{equation}
where $t$ is the transmittance of the beam splitter. 
Throughout the paper, we often simplify the description of 
a block diagonalized matrix like Eq.~(\ref{eq:beam_splitter}) as
\begin{equation}
S_{AB}^t
= \left[
\begin{array}{cc}
\sqrt{t} & \sqrt{1-t} \\
-\sqrt{1-t} & \sqrt{t}
\end{array}
\right]^{\oplus 2} . \nonumber
\end{equation}

\subsection{Measurement}
A major detection device in experimental photonic QIP
is a photon detector, which discriminates only zero or non-zero
photons. Such a device is mathematically described by a set of
measurement operators
\begin{equation}
\label{eq:PD_POVM}
\hat{\Pi}_{\rm off} = |0\rangle\langle0|, \quad
\hat{\Pi}_{\rm on} = \hat{I}-|0\rangle\langle0|,
\end{equation}
where $\hat{I}$ is an identity operator. 
Similar to the state, we can define the characteristic
function of the measurement operator $\hat{\Pi}$ as
$\chi_{\Pi} (x) = {\rm Tr}[ \hat{\Pi} \hat{\mathcal{W}}(x) ]$. 
In general, the probability of detecting the state $\hat{\rho}$ 
with the measurement operator $\hat{\Pi}$ is given by 
\begin{equation}
\label{eq:measurement}
{\rm Tr}\left[ \hat{\rho} \hat{\Pi} \right] 
= \left(\frac{1}{2\pi}\right)^n \int dx \, \chi_{\rho} (x) \chi_{\Pi} (-x) .
\end{equation}
Now suppose $\hat{\rho}$ is a single-mode Gaussian state 
with $\chi_\rho(x) = \exp[ -\frac{1}{4}x^T \gamma x ]$ and 
is measured by an on-off detector. 
The probability of obtaining the ``on'' outcome  (i.e. detecting 
non-zero photons) is calculated to be 
\begin{eqnarray}
\label{eq:photon_click}
P_{\rm on} & = & {\rm Tr}\left[ \hat{\rho} \hat{\Pi}_{\rm on} \right]
= 1 - {\rm Tr}\left[ \hat{\rho} |0\rangle\langle0| \right]
\nonumber\\
& = & 1-\frac{1}{2\pi} \int dx \, \chi_{\rho} (x) \chi_{\Pi_{\rm off}} (-x)
\nonumber\\
& = & 1-\frac{1}{2\pi} \int dx \, \exp \left[
-\frac{1}{4} x^T (\gamma + I) x \right]
\nonumber\\
& = & 1-\frac{2}{\sqrt{{\rm det}(\gamma+I)}} ,
\end{eqnarray}
where the last line performs the Gaussian integration. 
The above measurement is an ideal one, i.e. unit efficiency
and no dark counts. 
We will discuss the detector loss and dark counts 
in the next subsection.

In the SPDC QIP experiments, we often consider the coincidence
photon counts of a multi-mode quantum state.
Let $\hat{\rho}^\gamma$ be a density matrix of an $m$-mode Gaussian state
with covariance matrix $\gamma$. Then the following formula is useful
to calculate the coincidence counts:
\begin{eqnarray}
\label{eq:m_mode_no_click}
{\rm Tr}\left[ \hat{\rho}^\gamma |0\rangle\langle0|^{\otimes m} \right]
& = &
\left( \frac{1}{2\pi} \right)^m \int dx \, \exp\left[
-\frac{1}{4} x^T (\gamma+I) x \right]
\nonumber\\ & = &
\left( \frac{1}{2\pi} \right)^m \sqrt{
\frac{(4\pi)^{2m}}{\det (\gamma + I)} }
\nonumber\\ & = &
\frac{2^m}{\sqrt{\det(\gamma+I)}} .
\end{eqnarray}

\subsection{Imperfections}
{\bf Linear loss}. 
The optical channel with transmittance
$t$ (i.e. $1-t$ loss) is modeled by combining the channel with 
a vacuum environment via a beam splitter of transmittance $t$ 
and then tracing out the environment mode. 
The lossy channel is known as one of the Gaussian channels, i.e. 
it consists of Gaussian operations (but not necessarily unitary) 
\cite{Caruso2006}. 
Suppose a single-mode Gaussian state with covariance matrix $\gamma$ 
is transmitted through a lossy channel $\mathcal{L}^t$ where 
$t$ is the channel transmittance. 
Then $\mathcal{L}^t$ transforms the covariance matrix of the state as 
\begin{equation}
\label{eq:lossy_channel}
\mathcal{L}^t \gamma = K^T \gamma K + \alpha ,
\end{equation}
where $K = \sqrt{t} I$ and $\alpha = (1-t) I$. 
Note that the photon detector with efficiency $\eta$ 
can be modeled by a lossy channel with transmittance $\eta$ followed
by a lossless photon detector.

{\bf Detector dark counts}. 
The dark counts are modeled by Poissonian process \cite{Barnett1998} 
or phase insensitive amplification process \cite{Sekatski2012} 
those provide the same expression for the on-off detector operators: 
\begin{equation}
\label{eq:PD_POVM_DC}
\hat{\Pi}_{\rm off}(\nu) = (1-\nu)|0\rangle\langle0|, \quad
\hat{\Pi}_{\rm on}(\nu) = \hat{I}- \hat{\Pi}_{\rm off}(\nu) ,
\end{equation}
where $\nu$ is the dark count probability.

\begin{figure}
\begin{center}
\includegraphics[width=80mm]{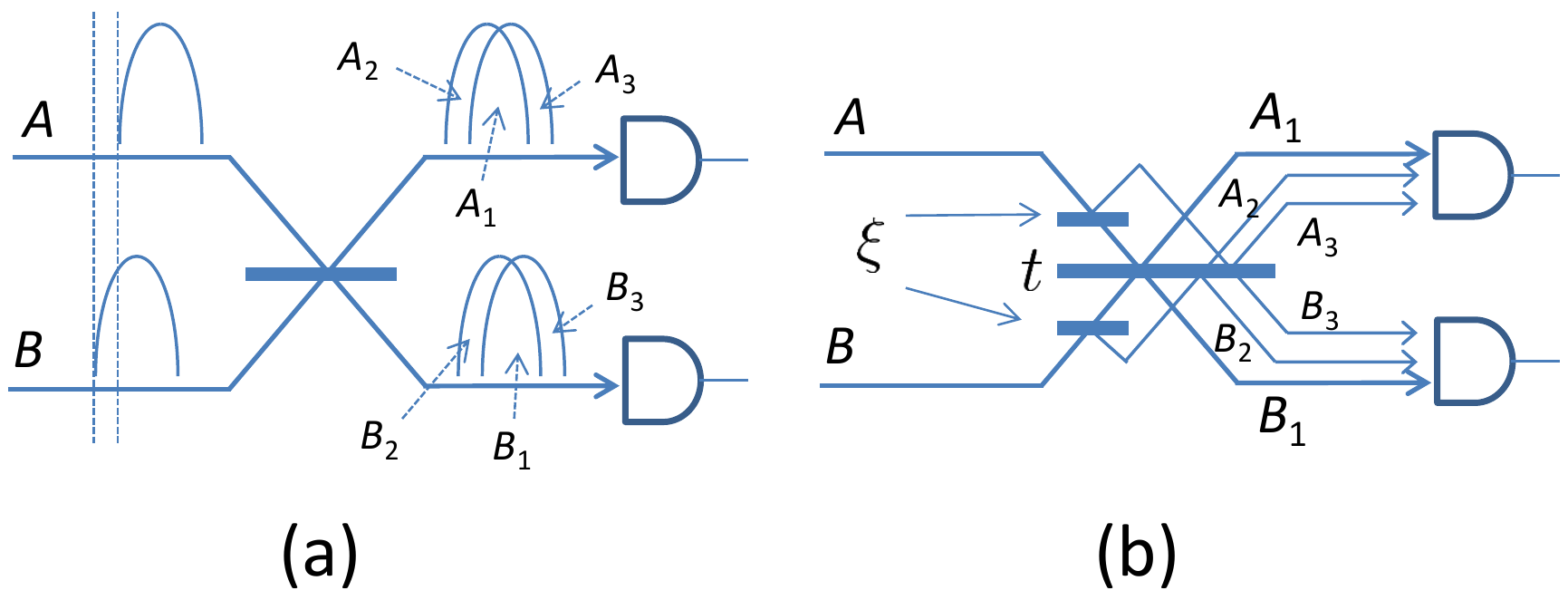}   %
\caption{\label{fig:mode_mismatch}
(a) Temporal mode mismatch at the beam splitter,
and (b) its model based on effective beam splitters.
$A_i$ and $B_i$ represent the modes where $A$ and $B$
corresponds to the spatial mode and the subscript
are the labels for the temporal modes.
Note that our model in (b) is phenomenological and thus
is not restricted to the temporal mode mismatch.
}
\end{center}
\end{figure}

{\bf Mode mismatch}. 
Imperfection of mode matching at a beam splitter
degrades the interference visibility.
Let $\xi$ be a phenomenological mode match factor 
representing the overlap between two pulses, i.e. 
$0\le\xi\le 1$ where $\xi=1$ corresponds to a perfect mode matching
while $\xi=0$ means the two pulses are in completely different modes 
and there is no interference between them. 
Then the effect of mode mismatch is modeled by
adding two (effective) beam splitters with transmittance $\xi$.
Figure \ref{fig:mode_mismatch}(a) and (b) depict
an example of the mode mismatch between two single-mode pulses 
in temporal mode and its translation into the beam splitting model 
(spatial mode), respectively, where the two pulses are partially overlapped. 
In spatial mode $A$ of Fig.~\ref{fig:mode_mismatch}(a), one can choose 
a temporal mode expansion in which one of the orthonormal mode 
is fully occupied by the pulse (i.e. includes $A_1$ and $A_3$). 
At the beam splitter, the mode expansion should be transformed 
such that the temporal part $A_1$ and $A_3$ are separated where 
these two modes should be a mixture with vacua to fulfil the normalization 
condition. This mode transformation is mathematically equivalent to 
split the pulse via a (virtual) beam splitter with transmittance $\xi$ 
into two spatial modes $A_1$ and $A_3$ (and combined it 
with a vacuum from the other port of the beam splitter). 
The same thing happens in spatial mode $B$ where the pulse is 
split into modes $B_1$ and $B_2$. Then $A_i$ and $B_i$ are 
combined at the real beam splitter where the pulses are interfered 
only between $A_1$ and $B_1$ (Fig.~\ref{fig:mode_mismatch}(b)).

Here we give an example. 
Suppose we would like to measure the coincidence counts
after interfering the two Gaussian state pulses via a beam splitter with
transmittance $t$. Let $\gamma_{A_1B_1}$ be the covariance
matrix of the two-mode state (i.e. two pulses) before the beam splitter.
Then after the beam splitting, a whole state is described by
a six-mode covariance matrix
\begin{eqnarray}
&& \tilde{\gamma}_{A_1B_1A_2B_2A_3B_3} = \nonumber\\
&& S^T_{\rm BS} S^T_{\rm MM}
(\gamma_{A_1B_1} \otimes I_{A_2B_2A_3B_3})
S_{\rm MM} S_{\rm BS} ,
\end{eqnarray}
where
\begin{eqnarray}
\label{eq:mode_match}
S_{\rm MM} & = & S^{\xi}_{B_1B_3} S^{\xi}_{A_1A_2} , \\
\label{eq:3BS}
S_{\rm BS} & = & S^t_{A_3B_3} S^t_{A_2B_2} S^t_{A_1B_1} ,
\end{eqnarray}
and the terms in the rhs are the beam splitter matrices 
defined in Eq.~(\ref{eq:beam_splitter}). 

The coincidence count probability is then given by
\begin{equation*}
{\rm Tr}[ \rho^{\tilde{\gamma}}
(\hat{I} - |0\rangle\langle0|^{\otimes 3})_{A_1A_2A_3}
(\hat{I} - |0\rangle\langle0|^{\otimes 3})_{B_1B_2B_3} ].
\end{equation*}
Note that in our model, the mode mismatch is included by
a phenomenological factor and thus we can incorporate
any kinds of mode mismatch, such as temporal, spectral, spatial, etc.

\subsection{Feedforward}
Before closing the section, we briefly mention the (classical) feedforward operations. Feedforward is an important resource in QIP. For example, it allows one to implement an on-demand single-photon source from heralding of the SPDC photons \cite{Migdall2002}, and even constructing a universal quantum computer \cite{Knill2001,Prevedel2007}. In the feedforward scenario, each operation in the system can be adaptively chosen according to the prior partial measurement outcomes. Therefore, to fully simulate such a system, one has to calculate all possible branches of the measurement outcomes and feedforwarded operations. As will be discussed in section 3.3, since the computational complexity of our method exponentially grows as the system size increases, with our method it is not possible to fully simulate a large scale linear optics quantum computer proposed in  \cite{Knill2001} with an efficient computation time. Note however, for relatively small size systems, it is possible to trace each feedforward branches and apply our method to calculate the probability of observing each event separately. That would be useful to simulate the currently (or near future) feasible experimental setups such as \cite{Migdall2002,Prevedel2007}.

\section{Applications}
\label{sec:applications}

In this section we apply our characteristic function approach
to SPDC based QIPs and compare some of them with 
previous experimental results. 

\subsection{Hong-Ou-Mandel interference of an SPDC source}

The Hong-Ou-Mandel (HOM) interference is observed when 
two indistinguishable single-photons are interfered via a 50/50 beam splitter 
\cite{Hong1987}.
The visibility of the HOM interference can be unit only when the two photons 
are fully indistinguishable in any degree of freedom (temporal mode, 
frequency mode, etc) which is thus a useful measure to evaluate 
the indistinguishability of the signal and idler photons 
from the same SPDC source. 
The HOM interference test for an SPDC source is described as follows. 
The signal and idler pulses are interfered
by a 50/50 beam splitter and the coincidence count of the two detectors
are measured by scanning the time delay of the idler pulse. 
The HOM dip is observed when there is no time delay for the idler pulse, 
i.e, the overlap of the signal and the idler in time is maximum.
Let $P^{CC}_{\rm min}$ be the coincidence count (CC) probability
without the delay and $P^{CC}_{\rm mean}$ be the CC probability
with the delay which is enough larger than the pulse width such that
no interference occurs between the signal and idler pulses.
Following the previous works e.g. \cite{Sekatski2012,Jin2014SR}, 
we define the visibility of the HOM test by 
\begin{equation}
\label{eq:HOM_visibility}
V_{\rm HOM} = \frac{P^{CC}_{\rm mean} - P^{CC}_{\rm min}}{P^{CC}_{\rm mean}}.
\end{equation}

\begin{figure}
\begin{center}
\includegraphics[width=80mm]{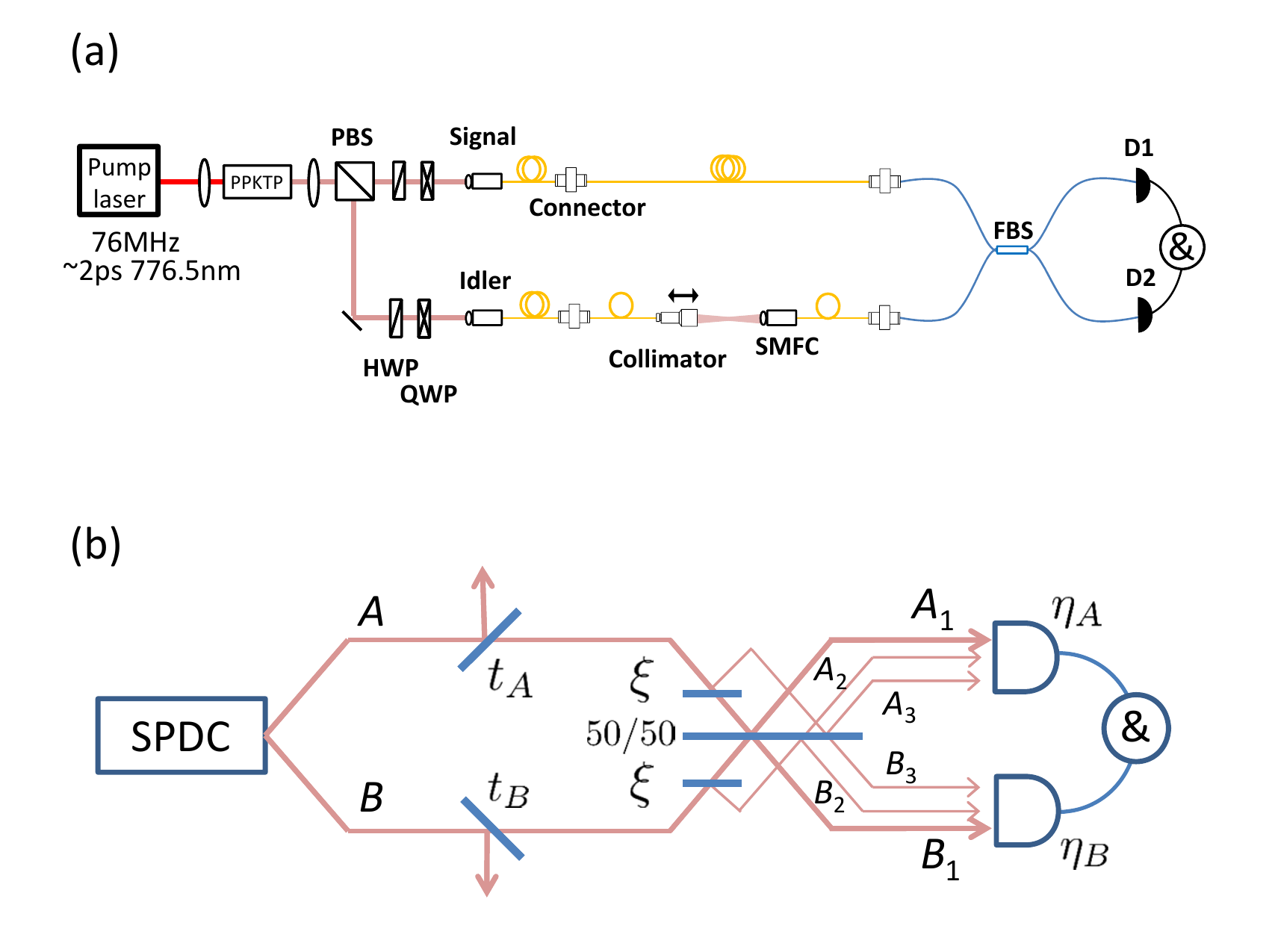}   %
\caption{\label{fig:HOM_setup}
(a) Experimental setup of the HOM experiment in \cite{Jin2014SR}, and
(b) the corresponding linear optics model.
PPKTP: periodically poled KTiOPO$_4$.
PBS: polarization beam splitter.
HWP: half waveplate.
QWP: quarter waveplate.
SMFC: single-mode fiber coupler.
FBS: fiber beam splitter.
}
\end{center}
\end{figure}

To see the validity of our method, we simulate the HOM experiment
in \cite{Jin2014SR} where the pump power dependence of $V_{\rm HOM}$
is experimentally observed with a standard mode-locked laser 
(Ti:Sapphire laser with a repetition rate of 76MHz).
The experimental setup and the corresponding theoretical model
are shown in Fig.~\ref{fig:HOM_setup}(a) and (b), respectively.

The output from the SPDC source is described by a TMSV
which fully includes the multi-photon components.
The transmission losses in the signal ($A$) and idler modes ($B$),
that are for example caused by coupling the spatial modes into fibers,
are described by beam splitters with the transmittance
$t_{A,B}$, respectively. 
Also $t_B$ includes the losses in the controllable delay line
(the collimator in Fig.~\ref{fig:HOM_setup}(a)) which makes the setup 
asymmetric (in the sense that $t_A \ne t_B$). 
$\eta_A$ and $\eta_B$ are the quantum efficiencies of the two photon
detectors. The mode mismatch between the signal and idler pulses
at the 50/50 beam splitter is characterized by the mode match factor $\xi$
which is in fact very sensitive to the HOM interference visibility
as shown below.

The covariance matrix of the state generated from the SPDC is 
given by $\gamma_{AB}^{\rm TMSV}(\mu)$ which is defined 
in Eq.~(\ref{eq:TMSV}). 
Applying losses in mode $A$ and $B$ with transmittance $t_A$ and $t_B$, 
respectively, we get 
\begin{eqnarray}
\label{eq:gamma_AB_after_loss}
\gamma^{\mathcal{L}}_{AB} & = & 
\mathcal{L}_B^{t_B} \mathcal{L}_A^{t_A} \gamma_{AB}^{\rm TMSV}(\mu)
\nonumber\\ & = & 
K_{AB}^{t_At_B \, T}  \gamma_{AB}^{\rm TMSV}(\mu) K_{AB}^{t_At_B}
+ \alpha_{AB}^{t_At_B}
\nonumber\\ & = & 
\left[
\begin{array}{cc}
2 t_A \mu + 1 & \pm 2\sqrt{t_A t_B \mu(\mu+1)} \\
\pm 2\sqrt{t_A t_B \mu(\mu+1)} & 2 t_B \mu + 1
\end{array}
\right]^{\oplus 2} ,
\nonumber\\
\end{eqnarray}
where 
\begin{equation}
\label{eq:K_tAtB}
K_{AB}^{t_At_B} = \left[
\begin{array}{cc}
\sqrt{t_A} & 0 \\
0 & \sqrt{t_B} 
\end{array}
\right]^{\oplus 2} ,
\end{equation}
and 
\begin{equation}
\label{eq:K_tAtB}
\alpha_{AB}^{t_At_B} = \left[
\begin{array}{cc}
1-t_A & 0 \\
0 & 1-t_B 
\end{array}
\right]^{\oplus 2} .
\end{equation}

Application of a 50/50 beam splitter with 
mode matching factor $\xi$ is calculated to be
\begin{eqnarray}
&& \gamma^{\rm BS}_{A_1A_2A_3B_1B_2B_3} 
\nonumber\\ &&
= S^T_{\rm BS} S^T_{\rm MM}
(\gamma^{\mathcal{L}}_{A_1B_1} \oplus I_{A_2B_2A_3B_3})
S_{\rm MM} S_{\rm BS} ,
\end{eqnarray}
where $S_{\rm MM}$ and $S_{\rm BS}$ are given by Eqs.~(\ref{eq:mode_match})
and (\ref{eq:3BS}) with $t=1/2$, respectively. 
$I_{A_2B_2A_3B_3}$ is a 8-by-8 identity matrix representing 
the vacua in modes $A_2$, $B_2$, $A_3$, and $B_3$. 
Finally photon detection consists of lossy channels corresponding
to the detector loss and perfect detectors.
The covariance matrix of the system after the lossy channels 
with $\eta_A$ and $\eta_B$, corresponding to the quantum efficiencies 
of two detectors, are given by 
\begin{equation}
\label{eq:gamma_HOM}
\gamma_{A_1 \dots B_3} = 
\mathcal{L}_{B_1B_2B_3}^{\eta_B} \mathcal{L}_{A_1A_2A_3}^{\eta_A} 
\gamma^{\rm BS}_{A_1 \dots B_3} .
\end{equation}

We are now at the position to derive the coincidence count probability:
\begin{widetext}
\begin{eqnarray}
\label{eq:P_CC_tilde}
P^{CC}_{\rm min} & = & {\rm Tr}\left[ \hat{\rho}^{\gamma_{A_1 \dots B_3}}
\left( \hat{I} - |0\rangle\langle0|^{\otimes 3} \right)_{A_1A_2A_3} 
\left( \hat{I} - |0\rangle\langle0|^{\otimes 3} \right)_{B_1B_2B_3}
\right]
\nonumber\\ & = &
1 - {\rm Tr}\left[\hat{\rho}^{\gamma_{A_1A_2A_3}}
|0\rangle\langle0|^{\otimes 3}_{A_1A_2A_3} \right]
- {\rm Tr}\left[\hat{\rho}^{\gamma_{B_1B_2B_3}}
|0\rangle\langle0|^{\otimes 3}_{B_1B_2B_3} \right]
+ {\rm Tr}\left[\hat{\rho}^{\gamma_{A_1 \dots B_3}}
|0\rangle\langle0|^{\otimes 6}_{A_1 \dots B_3} \right]
\nonumber\\ & = &
1 - \frac{8}{\sqrt{{\rm det}(\gamma_{A_1A_2A_3}+I)}}
- \frac{8}{\sqrt{{\rm det}(\gamma_{B_1B_2B_3}+I)}}
+ \frac{64}{\sqrt{{\rm det}(\gamma_{A_1 \dots B_3}+I)}} ,
\end{eqnarray}
where $\gamma_{A_1A_2A_3}$ and $\gamma_{B_1B_2B_3}$ are
the submatrices of $\gamma_{A_1 \dots B_3}$.
The determinants in Eq.~(\ref{eq:P_CC_tilde}) are explicitly given by
\begin{eqnarray}
\label{eq:detA_HOM}
\sqrt{{\rm det}(\gamma_{A_1A_2A_3}+I)}
& = &
8 \left\{
\left[ 1+ \frac{\eta_A \mu}{4} \left\{ 2(t_A + t_B)
- t_A t_B \eta_A (1-\xi^2)
\right\} \right]^2 - t_A t_B \eta_A^2 \xi^2 \mu (\mu+1) \right\}^{1/2},
\\
\sqrt{{\rm det}(\gamma_{B_1B_2B_3}+I)}
& = &
8 \left\{
\left[ 1+ \frac{\eta_B \mu}{4} \left\{ 2(t_A + t_B)
- t_A t_B \eta_B (1-\xi^2)
\right\} \right]^2 - t_A t_B \eta_B^2 \xi^2 \mu (\mu+1) \right\}^{1/2},
\\
\sqrt{{\rm det}(\gamma_{A_1 \dots B_3}+I)}
& = & 64 \left[ \left\{ 1+\frac{\mu}{4}\left[ 2(t_A + t_B) (\eta_A + \eta_B)
- t_A t_B \left\{(\eta_A + \eta_B)^2 - \xi^2 (\eta_A - \eta_B)^2 \right\}
\right] \right\}^2
\right. \nonumber\\ && \left.
- t_A t_B (\eta_A -\eta_B)^2 \xi^2 \mu (\mu+1)
\right]^{1/2} .
\end{eqnarray}
\end{widetext}
The coincidence count with delay, $P^{CC}_{\rm mean}$
is simply obtained from $P^{CC}_{\rm min}$ by setting $\xi = 0$, i.e.
no overlap between the two modes.

\begin{figure}
\begin{center}
\includegraphics[width=80mm]{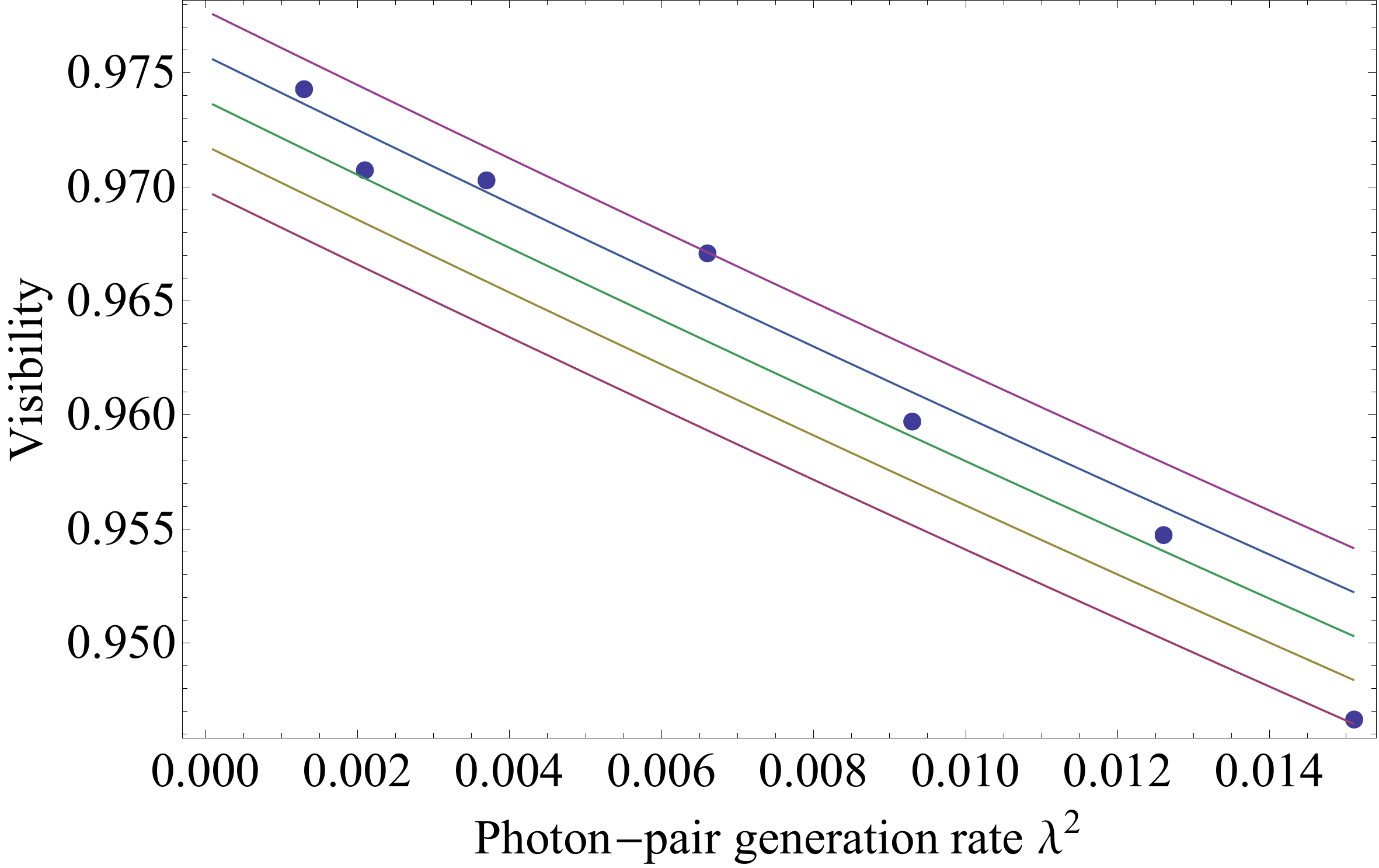}   %
\caption{\label{fig:HOM_experiment}
The experimental plots with the 76 MHz laser in \cite{Jin2014SR} and
the theoretical curves of the HOM visibility as a function of $\lambda^2$.
The theoretical curves are with $\xi=$0.9888, 0.9878, 0.9868, 0.9858, and
0.9849 from the top to the bottom.
}
\end{center}
\end{figure}

The theoretical visibilities derived above and
the experimental plots in \cite{Jin2014SR}
are compared in Fig.~\ref{fig:HOM_experiment} where
the two channel transmittances were experimentally measured as
$t_A = 0.42$, $t_B=0.29$. The detectors used in the experiment 
were superconducting nanowire single photon detectors (SNSPDs) that gave 
$\eta_A=0.68$, and $\eta_B=0.70$ and negligible dark counts 
(less than 1k counts per second). 
The squeezing parameter $\lambda^2$ corresponds to the photon-pair 
generation rate which is directly measurable in the experiment 
(recall that $\lambda^2 = \mu / (1+ \mu)$).
The mode matching factor is not measured and thus given
as a fitting parameter.
The figure shows a quantitative agreement between the theory 
and the experiment.
Also the theory curves reveal that the visibility is highly sensitive
to the mode matching at the beam splitter.
Note that \cite{Jin2014SR} also provides the theoretical curves, that are 
obtained by directly calculating the state vector of a whole system including
environments and numerically computed the coincidence counts.
One can numerically check that the numerical result in \cite{Jin2014SR} 
and the theoretical curve in Fig.~\ref{fig:HOM_experiment} 
are exactly identical. 
However, our method is much simpler and more systematic than the approach 
in \cite{Jin2014SR} 
and allows us to obtain a closed formula.
From that closed formula, we can also derive a simple analytical picture
of the visibility.
Suppose experimental imperfection is only
the low efficiency of the detectors $\eta_A=\eta_B=\eta \ll 1$
and $t_A=t_B=\xi=1$. Assuming $\mu \ll 1$, which is reasonable
for the photon-pair experiments, the visibility is simply given by
\begin{equation}
\label{eq:HOM_visibility_low_eta}
V_{\rm HOM} \approx \frac{1+2\mu}{1+4\mu} \approx 1 - 2\mu .
\end{equation}
It should be noted that Eq.~(\ref{eq:HOM_visibility_low_eta}) 
agrees with the one obtained in \cite{Sekatski2012}. 
Another interesting limit is 
the ideal case ($t_A=t_B=\eta_A=\eta_B=\xi=1$ and $\mu \ll 1$):
\begin{equation}
\label{eq:HOM_visibility_ideal}
V_{\rm HOM} \approx \frac{2+2\mu}{2+3\mu} \approx 1 - \frac{1}{2}\mu,
\end{equation}
which shows that the visibility degradation due to the multi-photon
emission is accelerated by the low efficiency of
the detectors.

\subsection{EPR interference of a Sagnac loop entanglement source}
\begin{figure}
\begin{center}
\includegraphics[width=80mm]{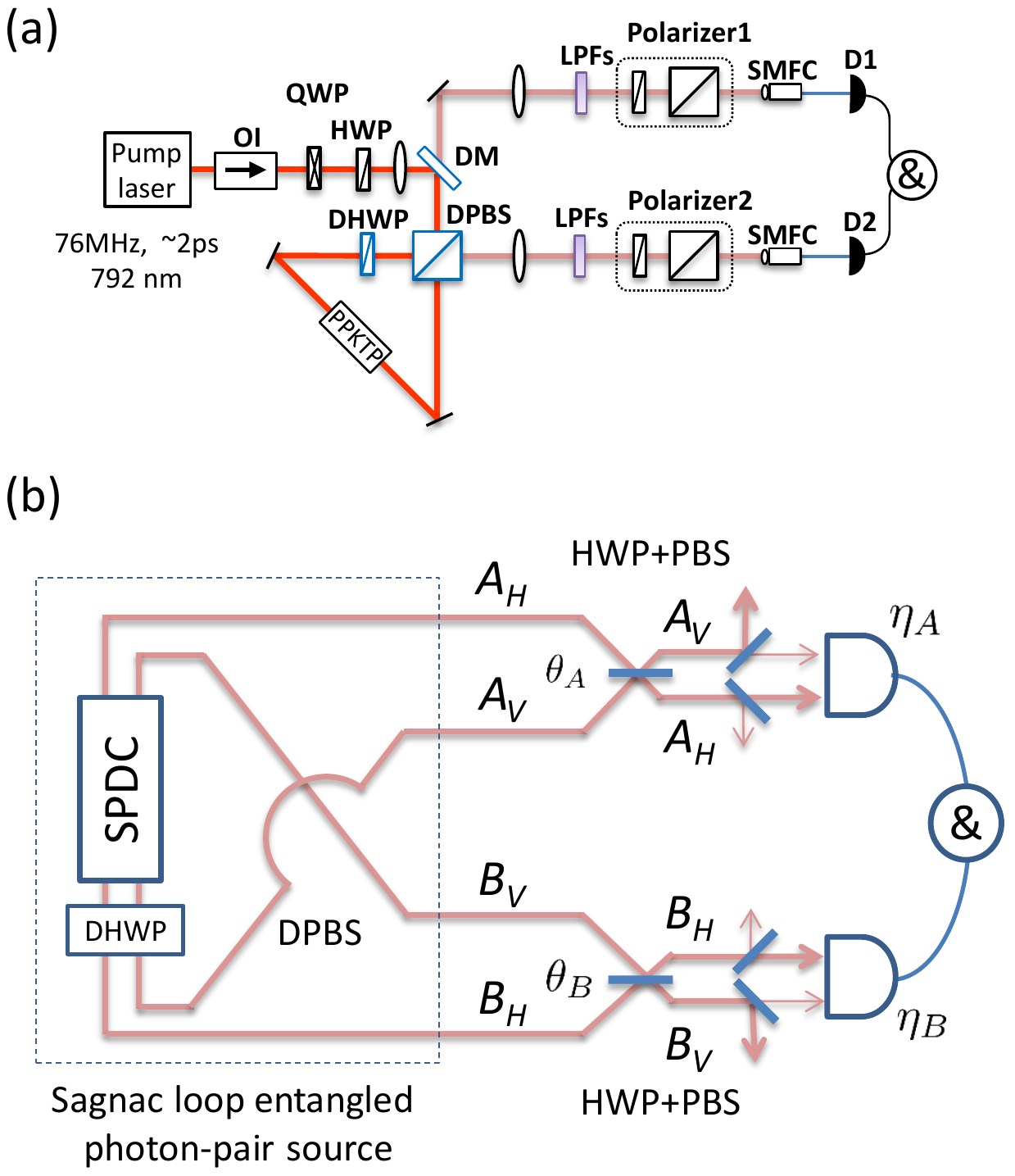}   %
\caption{\label{fig:EPR_setup}
(a) Experimental setup of the EPR interference experiment
in \cite{Jin2014OE}, and
(b) the corresponding linear optics model.
PPKTP: periodically poled KTiOPO$_4$.
OI: optical isolator.
HWP: half waveplate.
QWP: quarter waveplate.
DM: dichroic mirror. 
DPBS: dichroic polarization beam splitter.
DHWP: dichroic half waveplate.
LPF: long wavelength pass filter.
SMFC: single-mode fiber coupler.
}
\end{center}
\end{figure}

Another example is the EPR interference test (also called correlation 
measurement) for polarization entangled photon pairs.
We consider the Sagnac loop entanglement source, which was proposed 
in \cite{Shi2004} and is now widely used for photonic QIP experiments 
\cite{Jin2014OE,Kim2006, Fedrizzi2007, Prevedel2011, Jin2014swap}.
Here we compare our model with the EPR interference experiment performed
by a Sagnac loop entanglement source in \cite{Jin2014OE}.
The experimental set up and the corresponding theoretical model
are illustrated in Figs.~\ref{fig:EPR_setup}(a) and (b), respectively.

The Sagnac loop source generates a TMSV into the horizontal and
vertical polarization modes in the clockwise direction
(labeled as $A_H, A_V$ in the theoretical model) and the counter-clockwise
direction ($B_H, B_V$).
The entangled state is generated by swapping the vertical polarization modes
$A_V$ and $B_V$ via a dichroic polarization beam splitter (DPBS).
This operation transforms the covariance matrix of the two TMSV states
$\gamma_{A_HA_VB_HB_V}^{\rm TMSV2}$ into 
$\gamma_{A_HA_VB_HB_V}^{\rm SL}$ where
\begin{widetext}
\begin{eqnarray}
\label{eq:sagnac_loop_TMSV}
\gamma_{A_HA_VB_HB_V}^{\rm TMSV2} & = & \left[
\begin{array}{cccc}
2\mu + 1 & \pm 2\sqrt{\mu(\mu+1)} & 0 & 0 \\
\pm 2\sqrt{\mu(\mu+1)} & 2\mu + 1 & 0 & 0 \\
0 & 0 & 2\mu + 1 & \pm 2\sqrt{\mu(\mu+1)} \\
0 & 0 & \pm 2\sqrt{\mu(\mu+1)} & 2\mu + 1
\end{array}
\right]^{\oplus 2}, 
\\
\label{eq:sagnac_loop_entangled}
\gamma_{A_HA_VB_HB_V}^{\rm SL} & = & \left[
\begin{array}{cccc}
2\mu + 1 & 0 & 0  & \pm 2\sqrt{\mu(\mu+1)} \\
0 & 2\mu + 1 & \pm 2\sqrt{\mu(\mu+1)} & 0 \\
0 & \pm 2\sqrt{\mu(\mu+1)} & 2\mu + 1 & 0 \\
\pm 2\sqrt{\mu(\mu+1)} & 0 & 0 & 2\mu + 1
\end{array}
\right]^{\oplus 2} .
\end{eqnarray}
\end{widetext}

The EPR interference test is a common and relatively easy way to
experimentally estimate the quality of entanglement.
Each spatial mode is projected onto a particular polarization basis
by a half-wave plate (HWP) and a polarization
beam splitter (PBS) and then the photons in a chosen polarization are
detected by a single photon detector.
The coincidence photon count rate depends on the angle of each polarization
basis and the interference fringe is obtained by fixing one of
the polarization angles constant and rotate the other one.
The visibility is given by
\begin{equation}
\label{eq:entanglement_visibility}
V_{\rm ent} = \frac{P^{CC}_{\rm max} - P^{CC}_{\rm min}}{
P^{CC}_{\rm max} + P^{CC}_{\rm min}},
\end{equation}
where $P^{CC}_{\rm max}$ and $P^{CC}_{\rm min}$ are
the maximum and minimum count rates in the fringe.

\begin{figure}
\begin{center}
\includegraphics[width=80mm]{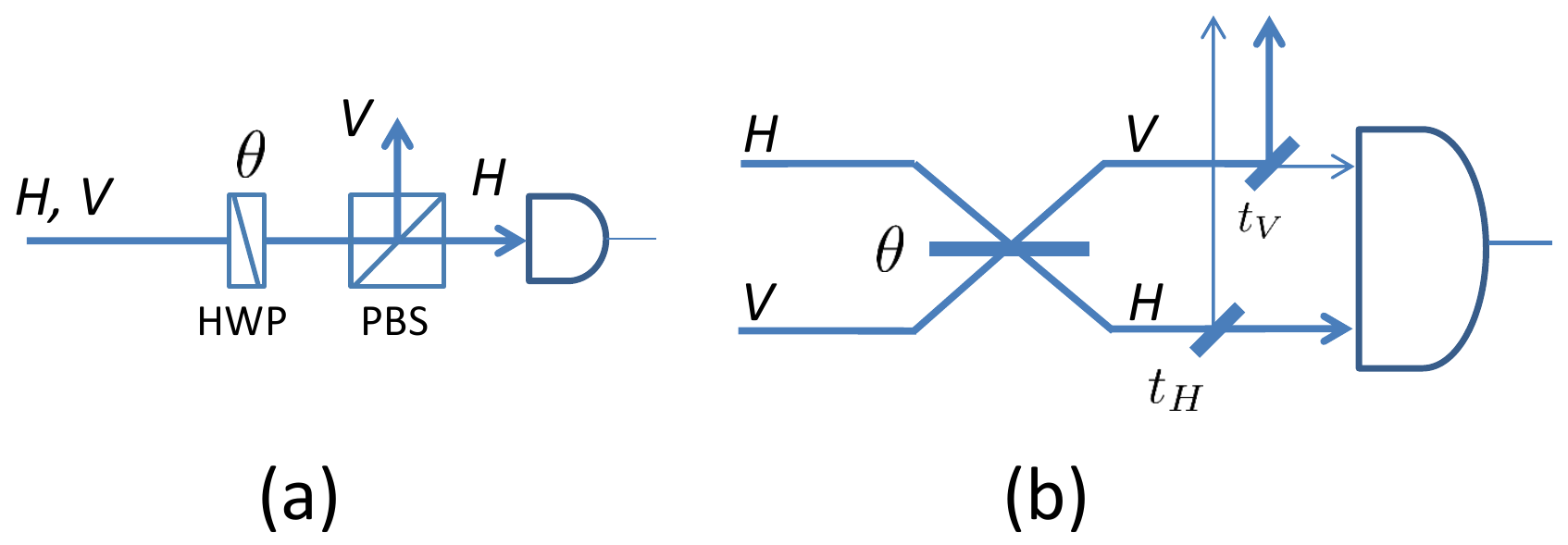}   %
\caption{\label{fig:HWP_PBS}
(a) The polarizer consisting of a half waveplate (HWP)
and a polarization beam splitter (PBS), and
(b) the corresponding linear optics model.
}
\end{center}
\end{figure}

The polarizer (HWP and PBS) with angle $\theta$ effectively works
as a beam splitter between the horizontal and vertical
polarization modes with transmittance $\cos^2\theta$. 
The main imperfection in this component
is a finite extinction ratio in the PBS which is modeled by
a perfect beam splitter followed by losses in each polarization mode
with $t_H < 1$ and $t_V > 0$ where 
$t_H$ and $t_V$ are the transmittances of the horizontal and
vertical polarization modes, respectively
(see Fig.~\ref{fig:HWP_PBS}(a) and (b)).
In the following, to simplify the calculation of the setup 
in Fig.~\ref{fig:EPR_setup}(b), we use 
$\tilde{\eta}_H = t_H \eta_H$ and
$\tilde{\eta}_V = t_V \eta_V$ where $\eta_H$ and $\eta_V$ are
the quantum efficiencies of the detector for horizontally and
vertically polarized photons, respectively.
The covariance matrix of the entangled source
in Eq.~(\ref{eq:sagnac_loop_entangled})
is first transformed by perfect beam splitter operations
$S^{\theta_B}_{B_HB_V} S^{\theta_A}_{A_HA_V}$. 
The imperfection of the PBSs and the imperfect quantum efficiency of the
detectors are included by applying the lossy channels 
$\mathcal{L}_{B_V}^{\tilde{\eta}_{B_V}} 
\mathcal{L}_{B_H}^{\tilde{\eta}_{B_H}}
\mathcal{L}_{A_V}^{\tilde{\eta}_{A_V}} 
\mathcal{L}_{A_H}^{\tilde{\eta}_{A_H}}$.
The coincidence count is then given by
\begin{widetext}
\begin{eqnarray}
\label{eq:P_CC_ent_vis}
P^{CC}_{\theta_A\theta_B} & = &
{\rm Tr}\left[ \hat{\rho}^{\gamma_{A_HA_VB_HB_V}}
\left( \hat{I} - |0\rangle\langle0|^{\otimes 2}
\right)_{A_HA_V}
\left( \hat{I} - |0\rangle\langle0|^{\otimes 2}
\right)_{B_HB_V}
\right]
\nonumber\\ & = &
1 - \frac{4}{\sqrt{{\rm det}(\gamma_{A_HA_V}+I)}}
- \frac{4}{\sqrt{{\rm det}(\gamma_{B_HB_V}+I)}}
+ \frac{16}{\sqrt{{\rm det}(\gamma_{A_HA_VB_HB_V}+I)}} ,
\end{eqnarray}
where
\begin{eqnarray}
\label{eq:detA_ent}
\sqrt{{\rm det}(\gamma_{A_HA_V}+I)}
& = &
4 (1+\tilde{\eta}_{A_H} \mu) (1+\tilde{\eta}_{A_V} \mu) ,
\\
\label{eq:detB_ent}
\sqrt{{\rm det}(\gamma_{B_HB_V}+I)}
& = &
4 (1+\tilde{\eta}_{B_H} \mu) (1+\tilde{\eta}_{B_V} \mu) ,
\\
\label{eq:detAB_ent}
\sqrt{{\rm det}(\gamma_{A_HA_VB_HB_V}+I)}
& = &
8 \left[ \left\{ 1+(\tilde{\eta}_{A_H}+\tilde{\eta}_{B_H}-\tilde{\eta}_{A_H}\tilde{\eta}_{B_H}) \mu
\right\} \left\{ 1+(\tilde{\eta}_{A_V}+\tilde{\eta}_{B_V}-\tilde{\eta}_{A_V}\tilde{\eta}_{B_V}) \mu
\right\} \right. \nonumber\\ && \left. +
\left\{ 1+(\tilde{\eta}_{A_H}+\tilde{\eta}_{B_V}-\tilde{\eta}_{A_H}\tilde{\eta}_{B_V}) \mu \right\}
\left\{ 1+(\tilde{\eta}_{A_V}+\tilde{\eta}_{B_H}-\tilde{\eta}_{A_V}\tilde{\eta}_{B_H}) \mu \right\}
\right. \nonumber\\ && \left. +
(\tilde{\eta}_{A_H}-\tilde{\eta}_{A_V})(\tilde{\eta}_{B_H}-\tilde{\eta}_{B_V}) \mu (\mu+1)
\cos 2(\theta_A+\theta_B) \right] .
\end{eqnarray}
\end{widetext}
The minimum and maximum count rates required 
in Eq.~(\ref{eq:entanglement_visibility}) are for example obtained by 
$P^{CC}_{\rm min}=P^{CC}_{0\,0}$ and $P^{CC}_{\rm max}=P^{CC}_{0 \, \pi/2}$ 
\cite{Jin2014OE}.

\begin{figure}
\begin{center}
\includegraphics[width=80mm]{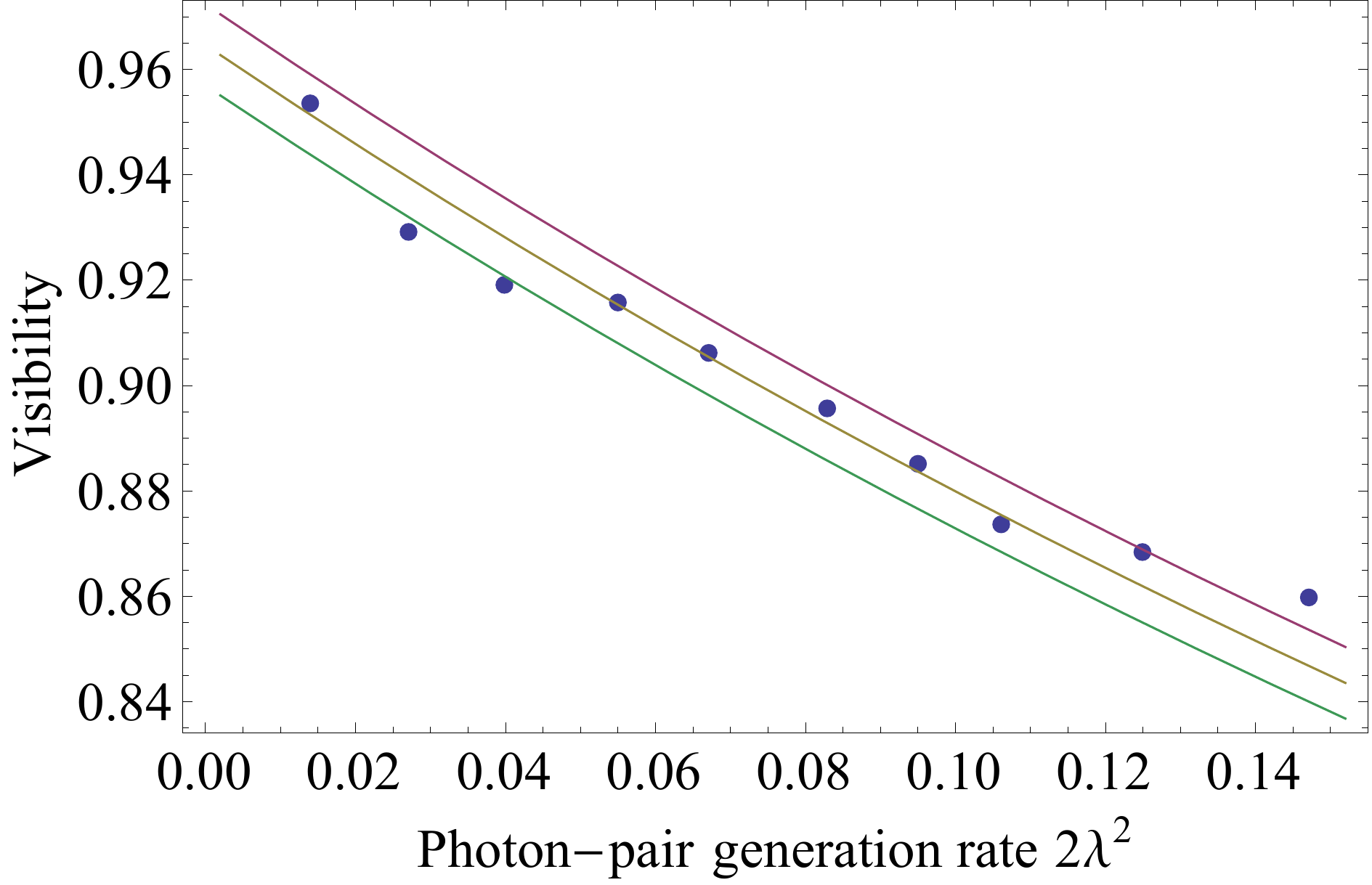}   %
\caption{\label{fig:EPR_experiment}
The experimental plots in \cite{Jin2014OE} and the theoretical curves
of the EPR interference as a function of the photon-pair generation 
rate of the Sagnac loop source $2\lambda^2$.
The theoretical curves are with $t_V=$0.007, 0.009, and 0.011
from the top to the bottom.
}
\end{center}
\end{figure}

Figure \ref{fig:EPR_experiment} shows the comparison with
the experiment in \cite{Jin2014OE} and our theoretical model. 
Note that the photon-pair generation rate of the Sagnac loop 
source should be $2 \lambda^2$ instead of $\lambda^2$ 
since the SPDC crystal is pumped twice (clockwise and counter-clockwise 
directions). 
In the experiment, the overall efficiencies for the horizontal
polarization are measured to be $\tilde{\eta}_{A_H} = \tilde{\eta}_{B_H}=0.1$
(again, SNSPDs are used as detectors and thus we neglect the effect of 
dark counts).
The transmittance of the vertical polarization at the PBS is not measured
and thus we use it as a fitting parameter varying from 0.007 to 0.011
($\le 0.01$ is guaranteed for a typical commercial PBS).
The theoretical estimate agrees with the experimental result.

Again, it is worth to further simplify the closed form given above.
By setting $\tilde{\eta}_{A_V} = \tilde{\eta}_{B_V} =0$ and 
$\tilde{\eta}_{A_H} = \tilde{\eta}_{B_H} = \eta$, the closed form 
of the visibility is reduced to 
\begin{equation}
\label{eq:ent_visibility_simple}
V_{\rm ent} = \frac{1+\mu}{1+3\mu+2\eta(2-\eta)\mu^2} .
\end{equation}
In the limit of weak pumping ($\mu \ll 1$), we have 
\begin{equation}
\label{eq:ent_visibility_low_eta}
V_{\rm ent} \approx 1 - 2\mu ,
\end{equation}
regardless of the detection efficiency $\eta$. 
Those results agree with the previous analyses 
in \cite{Kuzucu2008,Takesue2010,Sekatski2012}. 
\footnote{Note that the pump parameter $p$ in \cite{Sekatski2012} 
corresponds to our $\lambda^2$, i.e $p=\mu/(1+\mu)$.}.

\subsection{Concatenated entanglement swapping}
\label{subsec:CES}
The last application in this section is the concatenated 
entanglement swapping (CES) where we show how to apply our method to 
the QIP protocols with a multiple use of entangled source 
and demonstrate a drastic improvement of the computation time 
compared to the previous approach. 
Entanglement transmission over long distance is limited by 
the loss and noises in channel and detectors. 
Quantum repeater can overcome this limit but it is still challenging 
to implement it in large scale with the current technology. 
Concatenation of entanglement swapping (also called quantum relay) is known as 
another protocol to extend the distance of the entanglement distribution 
albeit with a resource overhead that is exponential in the distance (which is 
usually observed as an exponential decrease of the success probability) 
\cite{Gisin2002}. 
However, since entanglement swapping has been demonstrated with 
the current technology (see \cite{Jin2014swap} and references therein), 
it is still interesting to see the practical performance of the CES protocol 
in theory.

\begin{figure*}
\begin{center}
\includegraphics[width=160mm]{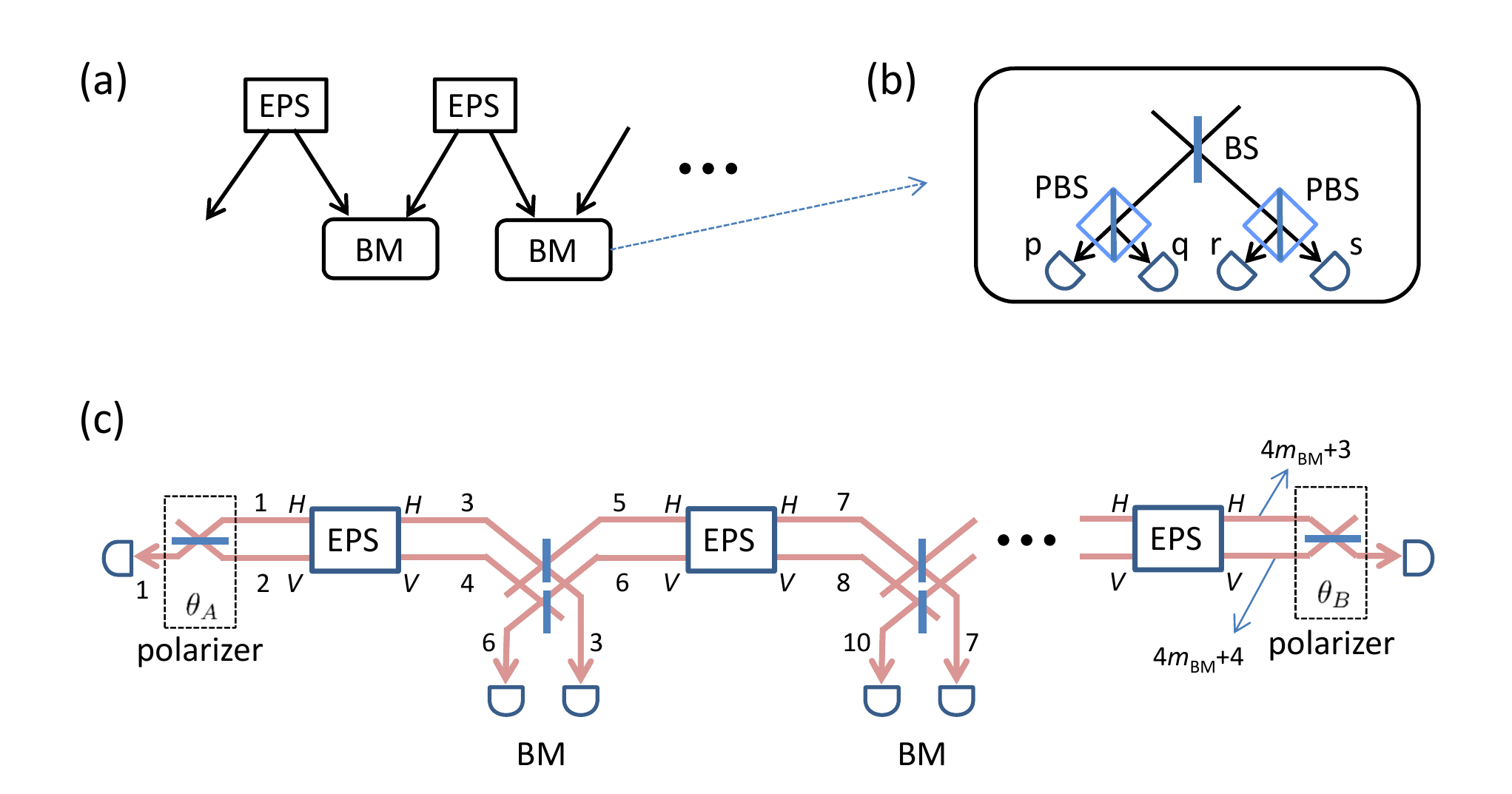}   %
\caption{\label{fig:CES_protocol}
(a) Schematic of the concatenated entanglement swapping and 
(b) the Bell measurement for polarized entangled photons. 
(c) Corresponding linear optics model (see the text for details). 
EPS: entangled photon source. BM: Bell measurement. BS: beam splitter. 
PBS: polarizing beam splitter. $m_{\rm BM}$: number of the Bell measurements 
used in the CES protocol. 
}
\end{center}
\end{figure*}

The protocol considered here is similar to the CES model 
discussed in detail in \cite{Khalique2013} where the state vector evolution 
of the two-mode squeezed state generated from SPDCs are calculated 
and the authors showed explicit expression of the density matrix 
and the detection probabilities those are used to derive the visibility 
of the EPR interference of the distributed entanglement numerically.  
Figure \ref{fig:CES_protocol}(a) illustrates the schematic of the protocol.
Entangled photon-pairs generated from the entangled photon-pair sources (EPSs) 
are swapped by the Bell measurement consisting of a 50/50 beam splitter, 
two PBSs and four on-off type photon detectors 
(Fig.~\ref{fig:CES_protocol}(b)). 
Entanglement swapping succeeds when photons are detected by 
particular two detectors in the Bell measurement 
for example at q and s in Fig.~\ref{fig:CES_protocol}(b). 
For more details of the protocol see \cite{Khalique2013} and 
the references therein.

Figure \ref{fig:CES_protocol}(c) is our model which is equivalent to 
Fig.~\ref{fig:CES_protocol}(a) but the orthogonal polarization modes are 
explicitly illustrated by different lines. 
For our EPS, we assume to use the Sagnac loop SPDC entangled source 
discussed in the previous subsection while 
the method can be applied to any other sources even not necessarily 
entangled in polarization modes (such as time-bin entanglements). 
For the system imperfections, we follow the assumptions used 
in \cite{Khalique2013}, i.e. losses in channels and detectors 
and the dark counts at detectors exist whereas the mode matching 
at beam splitters and the PBS devices are assumed to be perfect. 
The polarization modes are labeled by numbers from the left to the right 
(see Fig.~\ref{fig:CES_protocol}(c)). 
Also in the figure, for simplicity, only two detectors are illustrated 
for each Bell measurement where the simultaneous clicks 
at these detectors correspond to the successful CES. 
At the left and right ends, two polarizers (see Fig.~\ref{fig:HWP_PBS}) 
are placed to measure the EPR interference visibility of the swapped state. 

We now calculate the joint detection probability of all detectors 
illustrated in Fig.~\ref{fig:CES_protocol}(c) as a function of 
two polarizer angles $\theta_A$ and $\theta_B$. 
Let $m_{\rm BM}$ be the number of Bell measurements (the number of 
concatenation), i.e. there are $m_{\rm BM}+1$ Sagnac loop sources (EPSs).  
The covariance matrix for the quantum state generated from the most left 
Sagnac loop SPDC is $\gamma^{\rm SL}_{1234}$ as discussed 
in Eq.~(\ref{eq:sagnac_loop_entangled}). 
The total quantum state generated from $m_{\rm BM}+1$ Sagnac loops is thus 
simply given by $\gamma^{\rm SL \, \oplus (m_{\rm BM}+1)}$, more precisely,  
\begin{equation}
\label{eq:m+1SagnacLoops}
\left[
\begin{array}{ccc}
\gamma^{\rm SL}_\pm (\mu) & 0 & \cdots \\
0 & \gamma^{\rm SL}_\pm (\mu) &  \\
\vdots & & \ddots
\end{array}
\right]^{\oplus 2} ,
\end{equation}
where 
$\gamma^{\rm SL} \equiv \left( \gamma^{\rm SL}_\pm (\mu) \right)^{\oplus 2}$ 
and thus $\gamma^{\rm SL}_\pm (\mu)$ correspond to 
the two quadrature components of $\gamma^{\rm SL}$. 
Following the discussion in \cite{Scherer2009,Khalique2013}, 
all transmission loss is included in the detector efficiency. 
We also assume that all transmission channels have the same loss 
and all detectors are identical.
Therefore, we first apply the beam splitting operations 
at the Bell measurements. 
The beam splitters combine modes $4x-1$ and $4x+1$ 
for horizontal polarizations and modes $4x$ and $4x+2$ 
for vertical polarizations where $x=1, \dots, m_{\rm BM}$. 
Thus the symplectic matrix applied to 
$\gamma^{\rm SL \, \oplus (m_{\rm BM}+1)}$ is 
\begin{equation}
\label{eq:mBMBS}
S_{\rm BM} = \prod_{x=1}^{m_{\rm BM}} S^{1/2}_{4x-1,4x+1} S^{1/2}_{4x,4x+2}.
\end{equation}
The two polarizer operations at the end of the concatenation are also given by 
applying the beam splitter symplectic matrix 
\begin{equation}
S_{\rm AB} = S^{\theta_A}_{1,2} S^{\theta_B}_{4m_{\rm BM}+3,4m_{\rm BM}+4}.
\end{equation}
Let $\eta$ be the detector efficiency (including the channel transmittance). 
This is included by applying
\begin{equation}
\label{eq:total_loss}
\mathcal{L}^\eta_{\rm tot} = \mathcal{L}^\eta_{4m_{\rm BM}+4} \cdots 
\mathcal{L}^\eta_1 ,
\end{equation}
to the covariance matrix of the state. 
In total, we have the covariance matrix
\begin{equation}
\label{eq:cov_fin}
\gamma^{CES}_{m_{\rm BM}} = \mathcal{L^\eta_{\rm tot}} \left(
S_{\rm AB}^T S_{\rm BM}^T \gamma^{\rm SL \, \oplus (m_{\rm BM}+1)} 
S_{\rm BM} S_{\rm AB} \right) .
\end{equation}

The joint detection probability of all detectors illustrated in 
Fig.~\ref{fig:CES_protocol}(c) is now obtained by calculating 
\begin{widetext}
\begin{eqnarray}
\label{eq:CES_pcc}
P^{CES}_{\theta_A \theta_B} & = & 
{\rm Tr} \left[ \hat{\rho}^{\gamma^{CES}_{m_{\rm BM}}} 
\left( \hat{I}-(1-\nu)|0\rangle\langle0| \right)_1 
\left( \hat{I}-(1-\nu)|0\rangle\langle0| \right)_{4m_{\rm BM}+3} 
\prod_{x=1}^{m_{\rm BM}}  
\left( \hat{I}-(1-\nu)|0\rangle\langle0| \right)_{4x-1} 
\left( \hat{I}-(1-\nu)|0\rangle\langle0| \right)_{4x+2} \right]
\nonumber\\ & = & 
1 + \sum_{m=1}^M \sum_{C(M,m)}^{_M C_m} 
\frac{\{ -2(1-\nu) \}^m}{ 
\sqrt{ {\rm det}(\gamma^{\rm CES}_{C(M,m)}+I) } } ,
\end{eqnarray}
\end{widetext}
where $\nu$ is the dark count rate of detectors, 
$M$ is the total number of clicked detectors, i.e. $M=2m_{\rm BM}+2$, 
and $\gamma^{\rm CES}_{C(M,m)}$ is the submatrix of 
$\gamma^{\rm CES}$ 
taking only modes $C(M,m)$. 
$C(M,m)$ is the $m$-detector combination from $M$ detectors. 
For example, when $m_{\rm BM}=1$, total number of detectors is 
$M=4$ and the detectors are placed in modes 1, 3, 6, and 7. 
Then $C(4,1)=$1, 3, 6, 7, $C(4,2)=$13, 16, 17, 36, 37, 67, 
$C(4,3)=$136, 137, 167, 367, and $C(4,4)=$1367.
For reference, we describe a detailed structure of 
$\gamma^{CES}_{\rm m_{\rm BM}}$ in Appendix.

\begin{figure}
\begin{center}
\includegraphics[width=80mm]{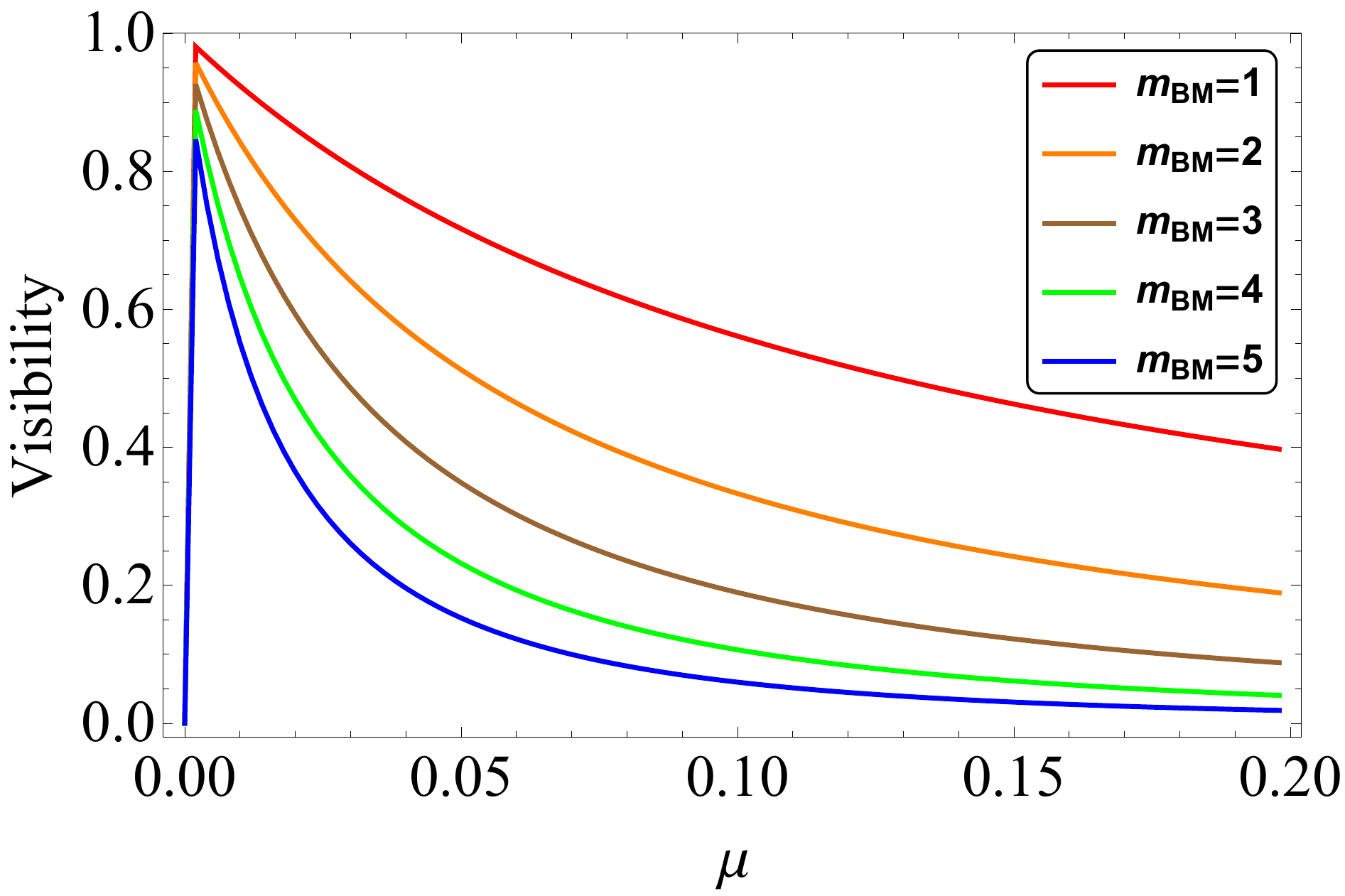}   %
\caption{\label{fig:CES_visibility1}
The EPR interference visibility of the entangled state 
distributed by the concatenated entanglement swapping 
as a function of the squeezing parameter $\mu$ 
with $\eta=0.04$ and $\nu=10^{-5}$ (see the main text). 
The number of the Bell measurements (concatenation) is 
$m_{\rm BM}=$1, 2, 3, 4, and 5 for 
red, orange, brown, green, and blue lines. 
}
\end{center}
\end{figure}

\begin{figure}
\begin{center}
\includegraphics[width=80mm]{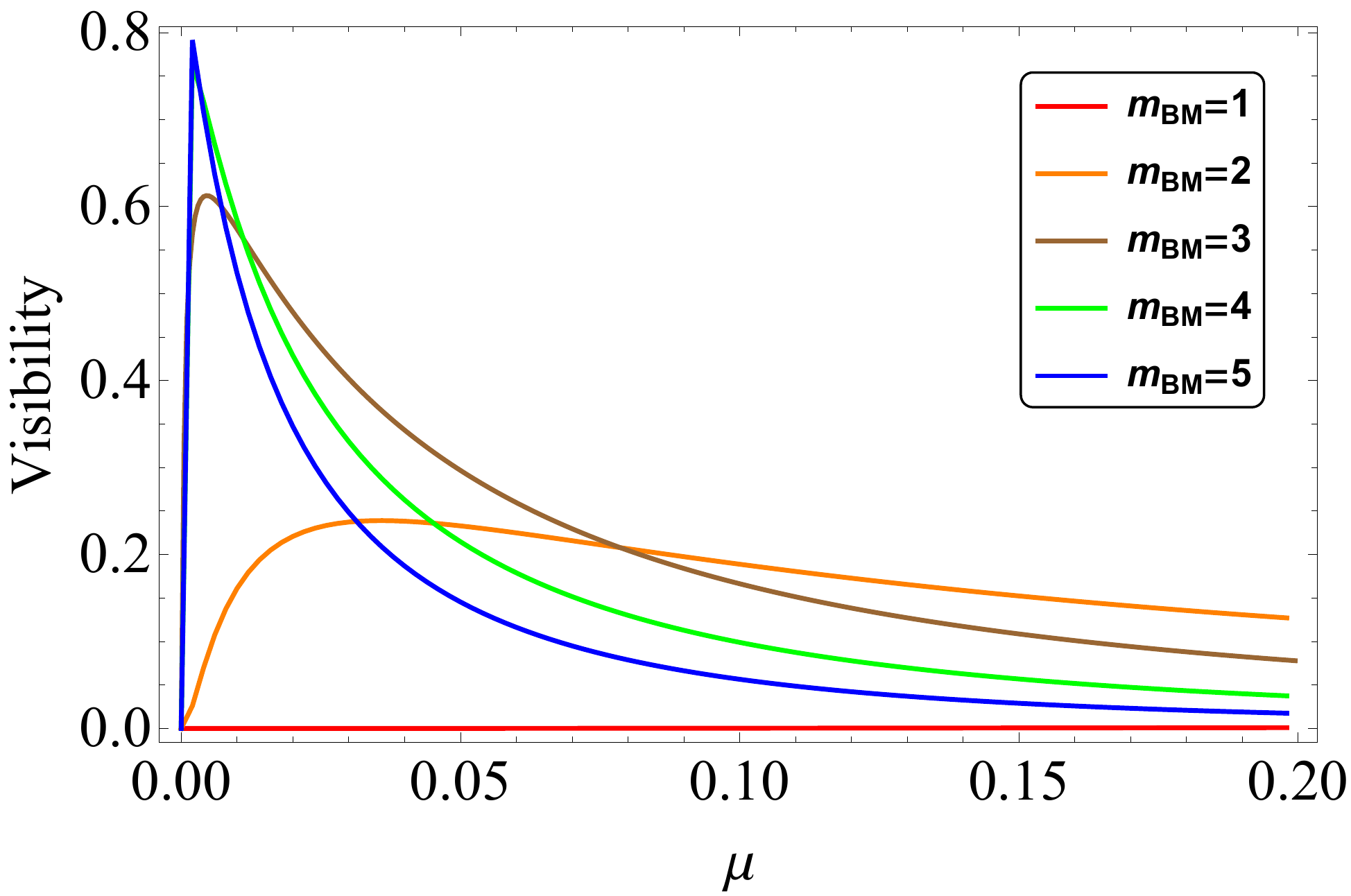}   %
\caption{\label{fig:CES_visibility2}
The EPR interference visibility for the same transmission 
distance with different number of concatenation. 
Transmittance of the whole channel is $10^{-20}$ which 
corresponds to 1000km distance by a standard fiber with 
loss of $0.2$dB/km. Efficiency and dark counts of the detectors 
are $\eta_{\rm D}=0.7$ and $\nu=10^{-5}$, respectively. 
The number of the Bell measurements (concatenation) is 
$m_{\rm BM}=$1, 2, 3, 4, and 5 for 
red, orange, brown, green, and blue lines. 
}
\end{center}
\end{figure}

In Fig.~\ref{fig:CES_visibility1}, the EPR interference visibilities 
are plotted as a function of $\mu$ for $m_{\rm BM}=1, \dots, 5$. 
For comparison, we choose the parameters used in \cite{Khalique2013}, 
$\eta=0.04$ and $\nu=10^{-5}$ for the transmittance and dark counts, 
respectively.  
In \cite{Khalique2013}, the CES with $m_{\rm BM}=3$ is numerically simulated 
by a super computer with the photon number truncation at 3 photons 
in each mode (the authors also reported that it takes 6 hours to get 
a single plot for the same simulation by a single-core use of 
a commercially available computer). 
In Fig.~\ref{fig:CES_visibility1}, the curve for $m_{\rm BM}=3$ (brown line) 
consists of 400 plots that are calculated by a commercial computer with 
only 10 seconds without any photon number 
truncation\footnote{Precisely, we use the Mathematica 9.0 program 
running at 2.9 GHz 
on an Intel Core i7-3520M dual-core processor with 8 GB of memory.}. 
Similarly, 100 plots for $m_{\rm BM}=5$ (blue line) require around 2 minutes 
with the same computer 
(it should be noted that our method works only for on-off detectors 
while the mathematical formula in \cite{Khalique2013} includes 
both for on-off and photon number resolving detectors). 

Our method is useful to estimate the performance of the CES 
for the long-distance entanglement transmission. 
In our model the total transmittance consists of the detector efficiency 
and the channel transmission as 
$\eta = \eta_{\rm D} \times 10^{- \alpha L /10}$ 
where $\eta_{\rm D}$ is the detector efficiency, $\alpha$ is the loss 
coefficient, and $L$ is the distance of the transmission channel. 
In the following, we consider the photon pairs at 1550 nm and use of 
a standard optical fiber with $\alpha=0.2$ as a transmission channel. 
Figure \ref{fig:CES_visibility2} compares the visibility for different 
$m_{\rm BM}$ where the total distance of the channel is fixed to be 1000 km. 
The detector efficiency and dark counts are assumed to be 
$\eta_{\rm D}=0.7$ and $\nu=10^{-5}$, respectively, could be typical 
parameters for SNSPDs discussed in previous subsections. 
For example, for $m_{\rm BM}=1$, the channel is divided into four arms 
and each arm has 250 km distance. 
In Fig.~\ref{fig:CES_visibility2}, the visibility is plotted for the CES 
up to $m_{\rm BM}=5$. First it shows that for $m_{\rm BM}=1$, it is 
completely impossible to distribute entanglement and even for $m_{\rm BM}=2$, 
the visibility must be around 0.2. 
Second, for extremely small $\mu$, the figure shows that 
$m_{\rm BM}$ should be large as much as possible while for the most 
region of $\mu$, the curve for $m_{\rm BM}=5$ decreases quickly 
as $\mu$ increases and $m_{\rm BM}=$3 or 4 are optimal for most of $\mu$. 
The result suggests that to construct a practical CES network, 
the optimal number of concatenation should be carefully chosen 
by considering a given distance (channel loss) and detector parameters.

Finally, it should be worth to mention about a general scaling of 
the computational complexity of our method with respect to the number of 
modes $N$ (or equivalently number of detectors to be detected). 
Our method is not efficient in the sense that the computational complexity 
grows exponential with respect to $N$. 
This is clearly observed in Eq.~(\ref{eq:CES_pcc}). 
There are two steps to compute Eq.~(\ref{eq:CES_pcc}). 
First, one has to calculate the covariance matrix including losses and 
beam splitters (in general any linear optics and even squeezing operations). 
This step consists of multiplications of square matrices and each of that 
is polynomial (at worst order of $O(N^3)$ by direct calculation). 
Second, the determinants of the (sub)matrices in each term of 
Eq.~(\ref{eq:CES_pcc}) should be calculated. 
While determinants can be calculated for example 
with $O(N^3)$ by the LU decomposition \cite{HornJohnson}, 
the number of terms in the second line of Eq.~(\ref{eq:CES_pcc}) 
is increase as $2^N$ since the first line 
of Eq.~(\ref{eq:CES_pcc}) includes 
$N$ multiplication of binomial terms 
$(\hat{I}-(1-\nu)|0\rangle\langle0|)$. 
This results the method inefficient (at least $O(2^N)$ for simulating 
large scale networks. 
Note that this is not only for the CES protocols but 
a general property for any linear optics networks with SPDCs and 
photon detectors. 
However, we should stress that, as shown in this subsection, 
for experimentally feasible size of the network with the current 
technology or even larger sizes, our method can work as a powerful tool 
to estimate the performance of the protocols 
with a practical computational time.

\section{Conclusion}
\label{sec:conclusion}

In summary, we have proposed a method to compute the SPDC based QIP
experiments theoretically, which fully involves the multi-photon emissions
and various experimental imperfections.
The key ingredient of our method is an application of the characteristic
function formalism which has been widely used in CV-QIPs,
in particular, for Gaussian states and operations.
We apply our methods to three examples, the HOM interference and 
the EPR interference experiments, and the concatenated entanglement 
swapping protocol. 
The first two examples are compared with the previously reported 
experimental results and show quantitative agreements. 
Moreover, we provide the analytical expression for the HOM and EPR 
visibilities that include full multi-photon components and various 
experimental imperfections. These could be useful for estimating 
the performance of various experimental setups.
In the third example, we numerically simulate the performance 
of the CES protocol up to five Bell measurements (concatenations) 
which requires only few minutes with a commercially available computer. 
Our method could be useful to estimate the practical performance 
of the SPDC based protocol with experimentally feasible or even 
larger size linear optics networks. 
Interesting future applications would include 
multi-partite entanglement generation \cite{Pan2012}, 
QKD \cite{Ekert1991} and quantum repeaters 
with SPDC sources such as heralded entanglements \cite{Barz2010,Usmani2012}.

Finally, a general computational complexity of our method with respect 
to the size of the system (mode numbers) is discussed. 
We show that the complexity is growing as $O(2^N)$ and thus inefficient 
for simulating very large scale systems beyond the current or near-future 
technologies. In general, linear optics network with perfect single photons 
and photon number resolving detectors, it is impossible to 
efficiently simulate the large system classically which is why 
the linear optics with feedback can construct a universal quantum 
computer \cite{Knill2001} (related to this, it is believed that 
other related computing ideas such as Boson Sampling \cite{Aaronson2010} 
cannot be efficiently simulated by classical computers). 
Although we feel this would also be the case for linear optics system 
with SPDCs and on-off detectors, it remains as an important open question 
whether there exists an efficient algorithm to simulate such a system. 
We stress that, however, our method is at least applicable to simulate 
the proof-of-principle demonstration of these protocols 
\cite{Prevedel2007,Broome2013,Spring2013,Tillmann2013,Crespi2013} and 
also estimating even larger system that could be a real target 
in near-future experiments.

\begin{acknowledgements}
We are grateful to Nicolas Sangouard, Witlef Wieczorek, and Harald Weinfurter 
for valuable suggestions. This work is supported by ImPACT Program of Council for Science, Technology and Innovation, Japan.
\end{acknowledgements}

\appendix*
\section{The covariance matrix in the CES protocol}
The explicit structure of $\gamma^{CES}(m_{\rm BM})$ in Eq.~(\ref{eq:cov_fin}) 
is 
\begin{equation}
\label{eq:gamma^CES}
\gamma^{CES}_{m_{\rm BM}} = \left[ 
\begin{array}{cccccc}
A_2 & \pm B^T & 0 & 0 & 0 & 0 \\
\pm B & A_4 & \pm C^T & 0 & 0 & 0 \\
0 & \pm C & A_4 & \pm C^T & 0 & 0 \\
0 & 0 & \pm C & \ddots & \pm C^T & 0 \\
0 & 0 & 0 & \pm C & A_4 & \pm D^T \\
0 & 0 & 0 & 0 & \pm D & A_2 
\end{array}
\right]^{\oplus 2} ,
\end{equation}
where 
\begin{eqnarray}
\label{eq:A-D}
A_2 & = & (2\eta\mu + 1) I_2 , \\
A_4 & = & \left[ 
\begin{array}{cc}
A_2 & 0 \\
0 & A_2 
\end{array}
\right] , \\
B & = & \eta \sqrt{2\mu (\mu+1)} \left[
\begin{array}{cc}
\sin\theta_A & \cos\theta_A \\
\cos\theta_A & -\sin\theta_A \\
-\sin\theta_A & -\cos\theta_A \\
-\cos\theta_A & \sin\theta_A 
\end{array}
\right] , \\
C & = & \eta \sqrt{\mu(\mu+1)} \left[ 
\begin{array}{cc}
J_2 & J_2 \\
-J_2 & -J_2 \\
\end{array}
\right] , \\
D & = & \eta \sqrt{2\mu (\mu+1)} \left[
\begin{array}{cccc}
\sin\theta_B & \cos\theta_B & \sin\theta_B & \cos\theta_B \\
\cos\theta_B & -\sin\theta_B & \cos\theta_B & -\sin\theta_B 
\end{array}
\right] , \nonumber\\ 
\end{eqnarray}
and 
\begin{equation}
\label{eq:I_2_J_2}
I_2 = \left[
\begin{array}{cc}
1 & 0 \\
0 & 1 
\end{array}
\right] , \quad
J_2 = \left[
\begin{array}{cc}
0 & 1 \\
1 & 0 
\end{array}
\right] .
\end{equation}



\end{document}